\documentclass{IEEEtran}
\usepackage{cite}
\usepackage[shortlabels]{enumitem}
\usepackage[super]{nth}
\usepackage{amsmath,amssymb,amsfonts}
\usepackage{graphicx}
\usepackage{textcomp,nicefrac}
\usepackage{xcolor}
\usepackage{soul}
\usepackage[normalem]{ulem}
\def\BibTeX{{\rm B\kern-.05em{\sc i\kern-.025em b}\kern-.08em
T\kern-.1667em\lower.7ex\hbox{E}\kern-.125emX}}
\begin{document}
\title{High-Level Software Tools for LLRF System Dedicated to Elliptical Cavities Management of European Spallation Source Facility }
\author{K. Klys, W. Cichalewski, W. Jalmuzna, A. Mielczarek, P. Perek, A. Napieralski, \IEEEmembership{Senior Member, IEEE}
\thanks{This work has been partially supported by the Polish Ministry of Science and Higher Education, decision number DIR/WK/2018/2020/02-2.

K. Klys, W. Cichalewski, W. Jalmuzna, A. Mielczarek, P. Perek, A. Napieralski, are with the Department of Microelectronics and Computer Science, Lodz University of Technology, 90-924 Lodz,
Poland (e-mail: kklys@dmcs.pl).}}

\maketitle

\begin{abstract}
The European Spallation Source (ESS) accelerator is composed of superconducting elliptical cavities. When the facility is running, the cavities are fed with electrical field from klystrons. Parameters of this field are monitored and controlled by the Low-Level Radio Frequency (LLRF) system. Its main goal is to keep the  amplitude  and  phase  at  a  given  set-point. The  LLRF  system  is  also responsible  for  the  reference  clock  distribution.

During  machine  operation  the  cavities  are  periodically  experiencing strain caused by the Lorentz force, appearing when the beam  is  passing  through  the  accelerating  structures. Even small  changes  of  the  physical  dimensions  of  the  cavity  cause  a shift of its resonance frequency. This phenomenon, called detuning, causes significant power losses. It is actively compensated by the  LLRF  control  system,  which  can  physically  tune  lengths  of the  accelerating  cavities with stepper motors (slow,  coarse  grained control) and piezoelements (active  compensation  during  operation  state). 

The paper describes implementation and tests of the software supporting  various  aspects  of  the  LLRF  system  and  cavities management. The Piezo Driver management and monitoring tool is  dedicated  for  piezo  controller  device.  The  LO  Distribution application is responsible for configuration of the local oscillator. The Cavity Simulator tool was designed to provide access to properties  of  the  hardware  device,  emulating  behaviour  of  elliptical cavities.  IPMI  Manager  software  was  implemented  to  monitor state of MicroTCA.4 crates, which are major part of the LLRF system architecture. All applications have been created using the Experimental  Physics  and  Industrial  Control  System  (EPICS) framework  and  built  in  ESS  EPICS  Environment  (E3).  

\end{abstract}


\section{Introduction}
\label{sec:introduction}
\IEEEPARstart{T}{HE} European Spallation Source (ESS) accelerator is under construction in Lund, Sweden. It is planned that it will be the most powerful source of neutrons based on the spallation phenomenon. The most crucial parts of the facility are protons accelerator and rotating tungsten target. The neutrons will be generated when protons, accelerated up to 2~GeV, hit the target. The resulting beams will be used in 16 research instruments for scattering measurements, spectroscopy, and diffraction. The facility will deliver exceptionally long pulses, as for spallation source, lasting 2.86~ms with a repetition rate of 14~Hz. This frequency is defined by the maximum speed of the rotating choppers \cite{b1}, \cite{b2}.

While protons are being accelerated, they are transported through various sections of the accelerator. The superconducting part will be composed of 120 niobium elliptical cavities and 30 spoke cavities. The elliptical resonators can be divided into two types: High-Beta and Medium-Beta. Both will operate at the frequency of 704.42 MHz. Low temperature of the cavities will be ensured by partial immersion in liquid helium cooled to 2 K\cite{b3}.

The energy required for proton acceleration, in the form of the electromagnetic field, is delivered by the Radio Frequency (RF) system. It converts energy from the AC network to RF frequencies with chain of conditioners, modulators and klystrons. The electric filed inside the cavity must have the accurate phase and amplitude to provide optimal energy transfer to the proton beam. Those parameters are controlled by the Low-Level Radio Frequency (LLRF) control system \cite{b4}.
 
\section{LLRF CONTROL SYSTEM}
\label{sec:llrf}
The main task of the LLRF system is to maintain stable amplitude and phase of the accelerating field, despite disturbances of external nature. The most important cause of the cavity detuning is its deformation under action of the Lorentz force. This is a consequence of significant gradient of the electric field (15~MV/m for Medium- and 18~MV/m for High-Beta) and the beam passing through the cavity. \cite{b4}, \cite{b5}.

\begin{figure}[h!]
\centerline{\includegraphics[width=3.5in]{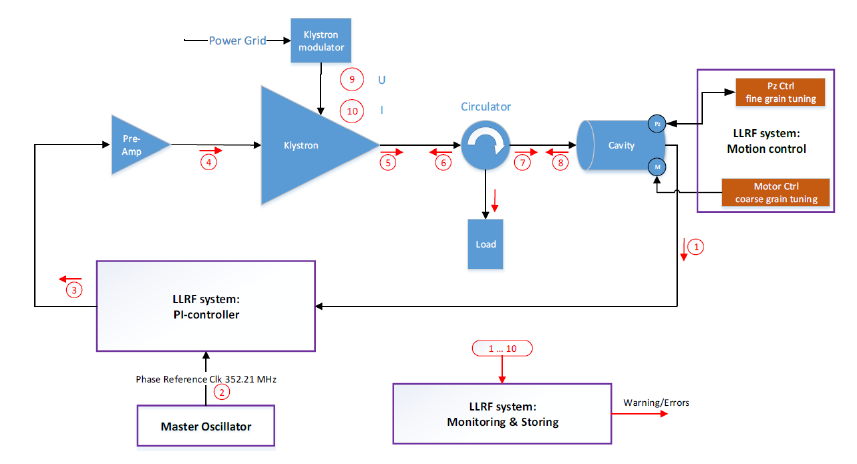}}
\caption{Schematic of LLRF section \cite{b5}}
\label{fig1}
\end{figure}

The control loop of the LLRF system is presented in \figurename~\ref{fig1}. The LLRF system measures electrical signals from the cavities that allow it to reconstruct the actual amplitude and phase of the field in the accelerating structure, in reference to the Master Oscillator reference clock. The calculated values are compared against the desired set-point using PI-controller. The controller computes the necessary phase shift and amplitude adjustment. These values are provided to a vector modulator which produces the corrected high frequency signal, using the Master Oscillator as a reference. Then the signal is boosted by a pre-amplifier and by a klystron -- powerful vacuum tube amplifier, with output power in the MW range. The cavity reflects some of the power delivered from klystron. A circulator is used to protect the vacuum tube from the returning power. Its task is to dump this wave to a dummy load, where it is converted to heat and dissipated \cite{b5}. 

Another task of the LLRF control system is physical tuning of the cavities. It manages two types of devices: stepper motors and dual piezo stacks. The former can stretch or squeeze the cavity during pre-tuning and operation. The later take part in fast tuning, which can be used when the accelerator is running. The piezoelectric stacks can operate as actuators or sensors. It means that they can be applied for measuring vibrations and stresses as well \cite{b6}.

The Master Oscillator is also a part of the LLRF system. It generates the phase reference clock and system clock to the whole accelerator. These signals provide reference timing for a wide variety of measurement and control systems interacting with the beam. 

The LLRF control system will be built using Micro Telecommunications Computing Architecture (MicroTCA.4) standard. This is a dedicated hardware platform for physics experiments, currently used in ITER tokamak and European X-ray Free-Electron Laser (E-XFEL) and many other large scale projects. Its main advantages are: high reliability, fair performance and easy manageability. This standard is derived from Advanced Telecommunication Computing Architecture (ATCA) and it implements the hot-swap technology.

Each slot of the MicroTCA.4 crate can host two units: an Advanced Mezzanine Carrier (AMC) in the front and an Rear Transition Module (RTM) in the back. The common practice in LLRF systems is to use the AMCs for digital processing, usually using FPGA circuits. The RTMs are, on the contrary, mainly used for analogue front-end, featuring down-converters, vector modulators, etc. \cite{b7}, \cite{b8}, \cite{b9}

\begin{figure}[h!]
\centerline{\includegraphics[width=3.5in]{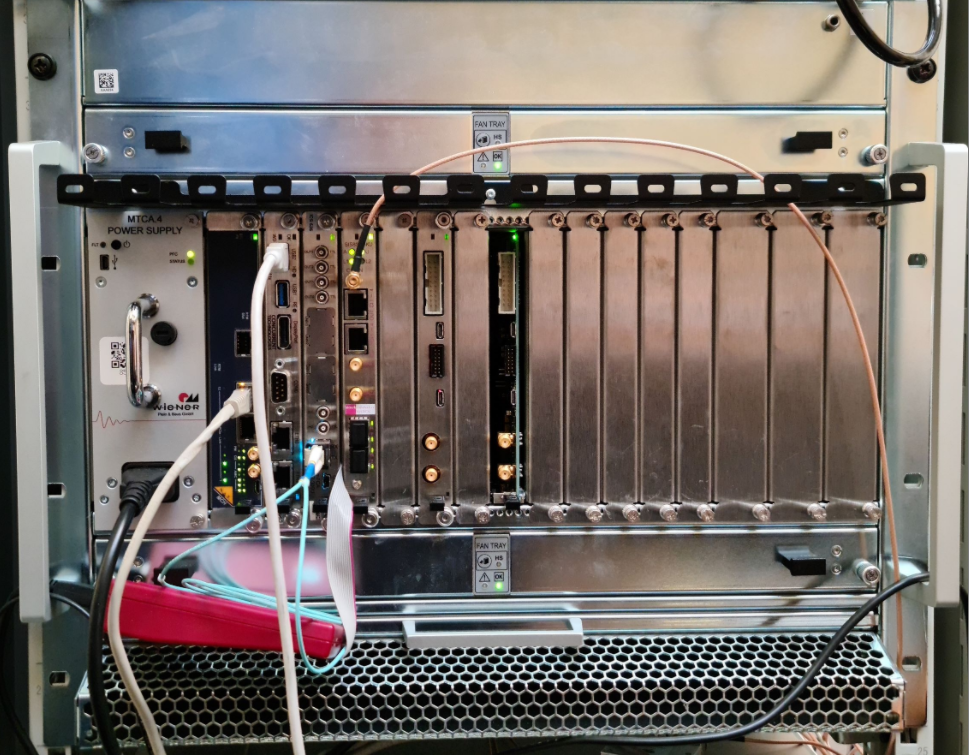}}
\caption{MTCA.4 crate for LLRF system}
\label{fig2}
\end{figure}

\section{EPICS and E3}
\label{sec:epics}

Experimental Physics and Industrial Control System (EPICS) is a framework devoted to design of distributed control systems. It is widely used in large research facilities like particle accelerators and telescopes. It allows for creation of servers, named IOCs (Input/Output Controller) which acquire data in real time (or perform other I/O activities) and then pass data via network protocols named Channel Access (CA) or PVAccess (PVA). The pieces of data are stored in Process Variables (PVs) which are organized in databases files \cite{b10}.

ESS EPICS Environment (E3) is an environment where EPICS software is compiled and launched. It was designed to keep consistency of EPICS applications. Each developer has its own way to develop IOCs and EPICS provides various possibilities to do this. Furthermore, each user can work with different hardware architecture on various software platforms. E3 was created to unify modules implementations. It keeps consistency and facilitates long-term maintenance. It lets user focuses on device integration, not on low-level development. Created IOC becomes a module which can be easily added to next applications. In the similar way EPICS modules like IocStats or Autosave are attached to developed IOC. Another advantage of the environment is the availability of templates and scripts, building empty E3 modules \cite{b11}. 

\section{Software tools}
\label{sec:tools}
The paper describes software created for elliptical cavities management and LLRF control system support:
\begin{enumerate}[(A)]
    \item Cavity Simulator management and monitoring software,
    \item Piezo Driver management and monitoring software,
    \item LO Distribution Tool,
    \item IPMI Manager supporting software.
\end{enumerate}

All the tools have been developed with EPICS framework built in E3 environment. The operator panels have been designed using CS-Studio Phoebus, ESS-site specific version. Some external action like saving data to file or loading them have been implemented using Jython scripts.

\subsection{Cavity Simulator Management and Monitoring Software}

\subsubsection{Cavity Simulator Design}

The Cavity Simulator was designed in order to minimise risk of damaging the cavities during field tests. It is used for development of LLRF software as well. The device can emulate the behaviour of the elliptical cavity of both types: Medium- and High-Beta. It reproduces phenomena like Lorentz force detuning, microphonics, beam loading and piezo compensation. The cavity model is represented as a parallel RLC circuit with the specific impedance. The resulting signals are calculated in Infinite Impulse Response (IIR) digital filter with adjustable parameters: quality factor, gain and detuning \cite{b12}, \cite{b13}.

\begin{figure}[h!]
\centerline{\includegraphics[width=3.5in]{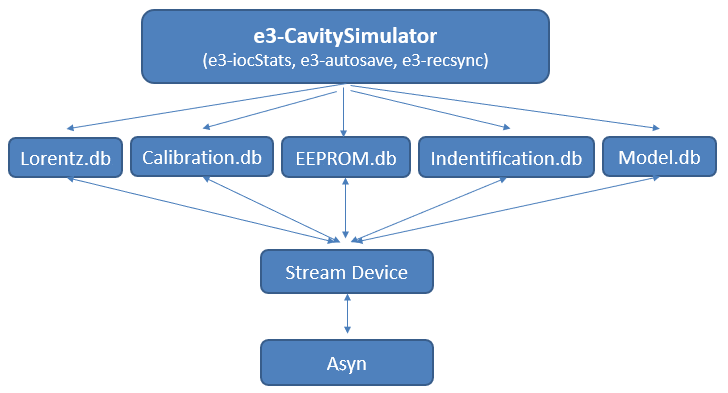}}
\caption{Simplified Cavity Simulator IOC schematic}
\label{fig3}
\end{figure}

The device contains high performance Field-Programmable Gate Array (FPGA) board, where simulation logic is stored, data converters, and custom RF front-end which is a bridge between LLRF control system and the Cavity Simulator. The simulator can be controlled remotely from an external PC through the Ethernet. Standard Commands for Programmable Instruments (SCPI) are used to communicate with the simulator and to change its configuration \cite{b12}, \cite{b13}. 

\subsubsection{Cavity Simulator Software Structure}

The designed E3 module uses Stream Device to send commands via Ethernet. This EPICS module is a device support, used for network communication. It demands the proper configuration like IP address and port, speed of the communication, stop and parity bits. In E3 environment, Stream Device is encapsulated into module that can be added to IOC in start command file. 

The software allows for configuration of all parameters of the simulator. Each database file corresponds to specific scope of device settings:
\begin{enumerate}
    \item Calibration.db - calibration of RF outputs,
    \item EEPROM.db - loading to EEPROM memory or saving to it,
    \item Identification.db - basic settings with software version readout,
    \item Lorentz.db - introducing Lorentz detuning force to cavity model,
    \item Model.db - cavity model parameters,
    \item Network.db - network settings of the device like IP address, port, DHCP,
    \item Piezo.db - piezo compensation effect, 
    \item SlowResonanceDrift.db - thermal turbulence's drift,
    \item Status.db - state of RF outputs,
    \item Temperature.db - temperature of sensors inside the simulator.
\end{enumerate}
In any database, PVs are divided into two groups: readouts and set-points. 

\subsubsection{Cavity Simulator IOC Operation}

The Graphical User Interface (GUI) is organised in a similar way as database files. Each window is dedicated to one set of parameters. The operator panels start with "Main menu" window, from where the user can access other screens. 

One of the most useful sections is "Model window" where the parameters of the cavity model can be modified. It provides the possibility of loading them from text file, as well as saving the current configuration. Another example is "Status window" with LEDs, informing about the state of each RF output and type of signal which is set. The settings can be checked and managed with combo-boxes and buttons. 

\subsection{Piezo Driver Management and Monitoring Software}

\subsubsection{Piezo Driver Design}

As mentioned earlier, the LLRF control system is based on MicroTCA.4 architecture. To ease integration and maintenance the Piezo Compensation System (PCS) was designed in the same form-factor. The PCS is composed of the hardware component, the Piezo Control Device (PCD), and the software support running on the CPU module. The PCD is designed as an RTM -- a card that is installed at the rear side of the control chassis. It is hosted by a dedicated carrier board, implemented as an AMC module equipped with a Xilinx Artix 7 FPGA. The PCS can share the crate with the LLRF system and even use the same CPU~\cite{b14}.

In the ESS accelerator the piezoelectric actuators will have to perform the compensation in a relatively short time (focused around RF pulses, lasting around 3 ms). This implies that a significant pulsed power has to be delivered to an actuator, in the range of up to 100 W per channel. There are no commercially available piezo-element drivers which are able to meet all of the ESS requirements. Therefore, a novel solution based on Class-D amplifiers was proposed \cite{b15}.

The main goal of the PCS is to drive the piezoelectric actuators according to pre-loaded waveforms, calculated by the control algorithm. At the same time it has to protect the piezo-element from excessive voltage, dissipated power and mechanical stress. A dedicated diagnostic microcontroller monitors the output current and voltage in the real time and can disable the power amplifier upon detecting dangerous driving signal. Finally, the PCS can also use the piezo-element as a strain sensor, by observing the voltage induced by the forces acting on the stack.

\begin{figure}[h!]
\centerline{\includegraphics[width=3.5in]{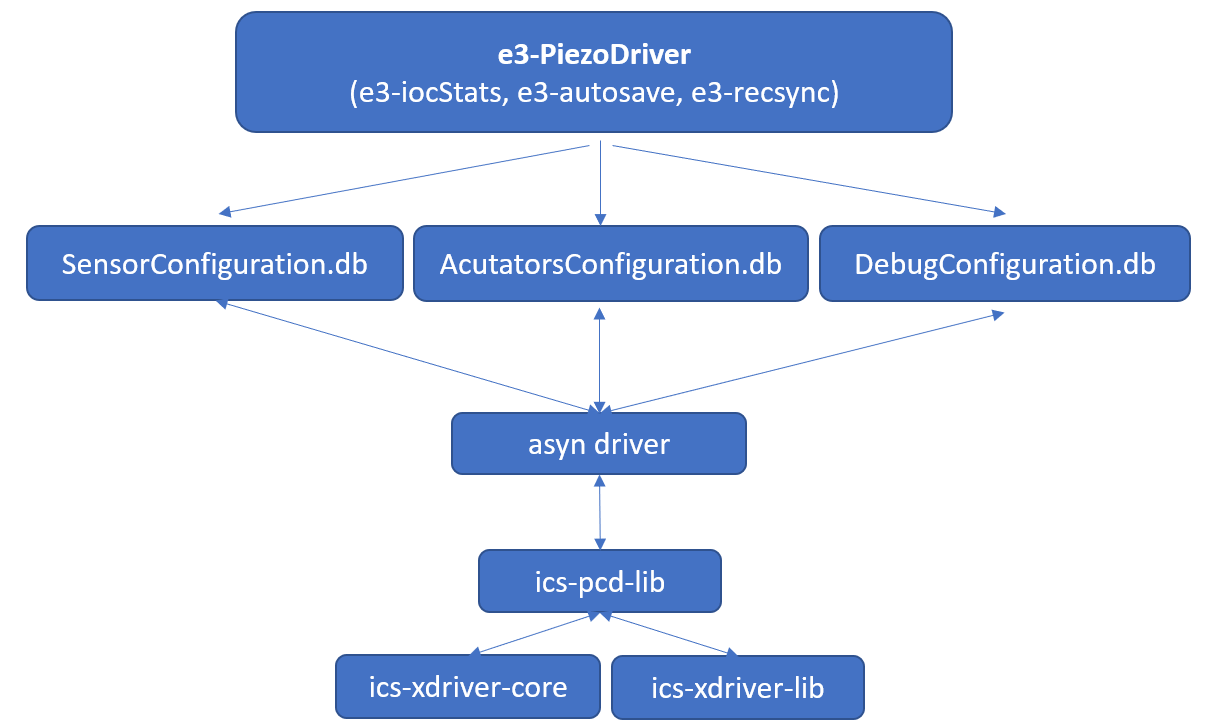}}
\caption{Piezo Driver IOC structure}
\label{fig4}
\end{figure}

The PCS should enable effective compensation of the Lorentz force detuning and microphonics. The driving signal for the actuators will be computed using detuning value from previous machine pulse, calculated by the LLRF system. There are various possible compensation methods investigated at the moment. Among them: single- and multi-pulse solutions are being evaluated. \cite{b13}, \cite{b15}. 

\subsubsection{Piezo Driver Software Structure}

The IOC structure was described in \figurename~\ref{fig3}. Designed E3 module contains other tools like IocStats or Autosave. The database files correspond to Piezo Driver functions: one file for sensor mode, second for actuators and last for general settings and mode changes. In order to use dedicated drivers, asyn C++ class was implemented. It uses Application Programming Interface (API) from ics-pcd-lib to communicate with the device. API provides C functions to control and read state of FPGA registers. Asyn class can be described as bridge between PVs and device. Thanks to asyn parameters that are assigned to each record, code can access and synchronize data.

When IOC is launched, the C++ constructor opens the communication band and it starts separate EPICS thread. This thread is responsible for acquiring data (it waits for flag informing whether new data are available and loads them to waveform PVs), synchronising data from/to PVs and for sending excitation signal to piezo elements.

Described API and drivers library are added to E3 module as Git (version-control system) sub-modules. When building IOC, they are downloaded from the repositories. Then, when the module is installed, the Dynamic Kernel Module Support (DKMS) adds kernel module for hardware driver and provides necessary libraries.

\begin{figure}[h!]
\centerline{\includegraphics[width=3.5in]{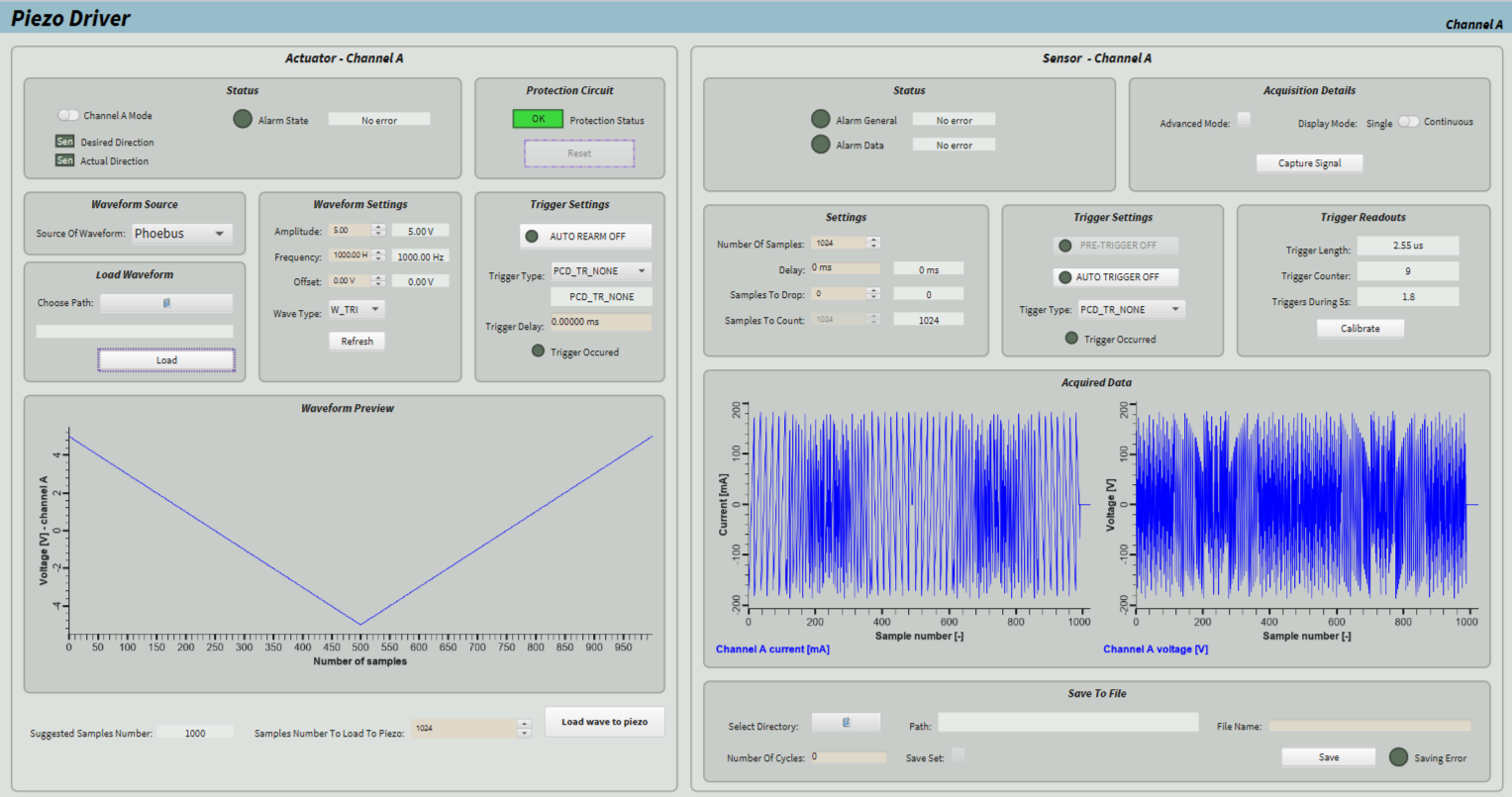}}
\caption{Piezo Driver operator panel}
\label{fig5}
\end{figure}

\subsubsection{Piezo Driver IOC Operation}

\figurename~\ref{fig4} shows operator panel for channel A (one of the piezo stacks). Left area provides overview of actuators settings. The excitation signal can be generated using predefined shapes and parameters, like amplitude, frequency, and offset. Its preview is visible below configuration fields. The operator can also load waveform from a given file. Right section is for sensor properties. There are two plots. First, representing temporal plot of the output current value. Second, providing waveform of voltage measured at the piezo stack terminals. They can be saved to file as well.

\subsection{LO Distribution Tool}

\subsubsection{LO Distribution Design}

The ESS LLRF control system masters clock signal distribution to other section of the accelerators. It also controls phase reference clock. The LO RTM device is responsible for the generation of the local oscillator signal used for RF signal down-conversion. It also produces sampling clock for ADC signal.

To be compatible with a MTCA.4, LO device was designed as RTM module. It is controlled via digital interface over the Zone3 connector by the AMC module \cite{b16}.

\begin{figure}[h!]
\centerline{\includegraphics[width=2.5in]{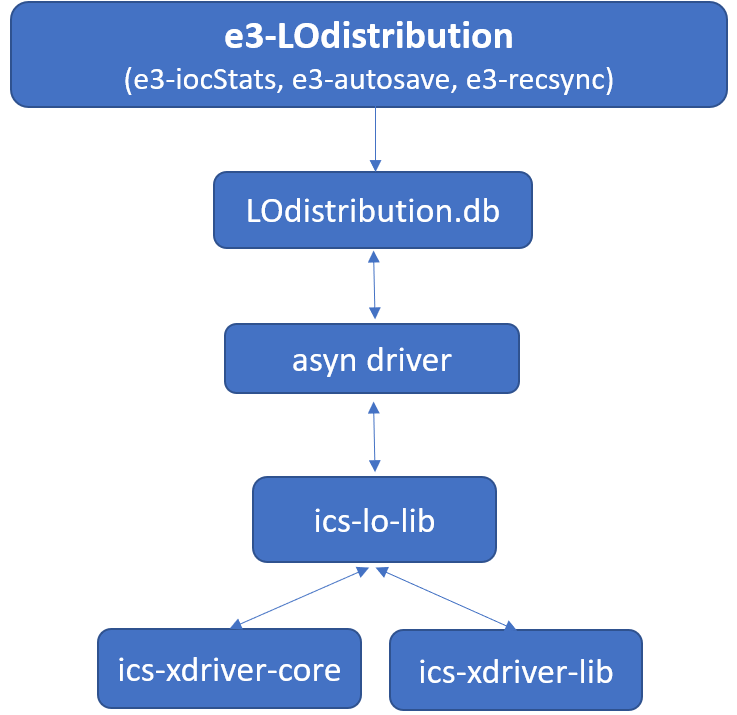}}
\caption{Structure of E3 LO module}
\label{fig6}
\end{figure}

\subsubsection{LO Distribution Software Structure}

The LO firmware is based on the same drivers as Piezo Driver (since they are both using same type AMC module – RTM carrier) but the different API library exposes unit parameters for EPICS IOC. The structure of E3 module is presented in \figurename~\ref{fig5}. The IOC structure is similar to Piezo Driver. Asyn C++ constructors initialize communication and the device itself. Then, in the newly created thread, IOC is synchronising values from database with data received from the device via API. Again, while E3 module is being built, kernel module is added to tree automatically. The drivers and API are attached to IOC as external Git sub-modules.

\subsubsection{LO Distribution IOC Operation}

The operator panels (\figurename~\ref{fig11}) provide set of buttons, sliders enabling LO configuration. An operator can decide when to activate LO and clock signals. The source of power trigger can be modified there. Furthermore, it is possible to modify properties of the LO signal like its frequency ratio, and boost type. Power readouts are available in dBm and in hex. they can be used to verify if the LO RTM is functioning properly. 

\subsection{IPMI Manager Supporting Software}

Since major part of the LLRF control system is MTCA.4 crates based systems it is important to provide methods to continuously monitor state and operational parameters of functional modules. The management layer of MTCA standard is based on Intelligent Platform Management Interface (IPMI). This protocol is used for intelligent monitoring of computer and telecommunications systems. Then, it was adapted to supervise MTCA, AMC, and ATCA standards \cite{b17}.

\subsubsection{IPMI Manager Design}

The implemented tool should be scalable. Since in the development phase, it can be tested only on several creates systems? with limited module configurations, it is crucial to provide a way to quickly adapt software for new MTCA crate. Together with ESS, it has been decided that the software should be integrated with EPICS framework. The following requirements for the tool have been defined:
\begin{enumerate}
    \item reading M-states of modules (M0-M7),
    \item reading identification information of each module,
    \item reading current values of various sensors, together with their threshold,
    \item readout of e-keying information,
    \item setting thresholds for each sensor,
    \item monitoring of MTCA crate events.
\end{enumerate}

The functional tests of two libraries, "OpenIPMI" and "openHPI" have been performed \cite{b18}, \cite{b19}. It was crucial to conduct tries using real crates because MTCA.4 standards introduce some changes to management layer. Both libraries provided defined functions however, "openHPI" is designed for monitoring of various IPMI systems. This leads to interface unification. Moreover, some information are hidden. In terms of EPICS software development it can cause different problems (for instance lack of access to detailed information about each module or sensor). Due to the reasons, we have? decided to use "OpenIPMI" library \cite{b18}.

\begin{figure}[h!]
\centerline{\includegraphics[width=2.5in]{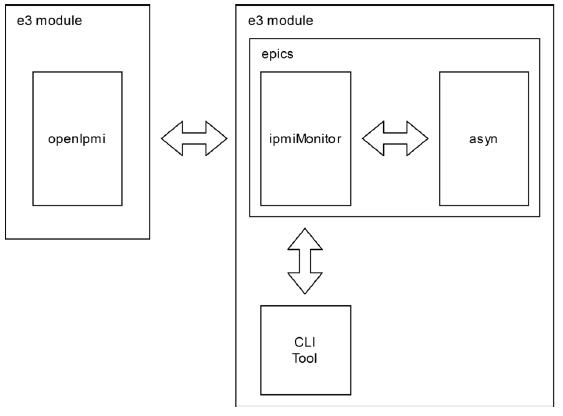}}
\caption{IPMI Manager architecture}
\label{fig7}
\end{figure}

\subsubsection{IPMI Manager Software Structure}

The structure of the IOC is presented in the \figurename~\ref{fig6}. It has been implemented as two modules. First one encapsulates the whole "OpenIPMI" into E3 module, to keep the library sources as a separate unit. The library itself has a form of independent module with C callback interface. In order to disable some ATCA function present in the library, the dedicated patch is applied at build time. There is also plugin for handling MTCA features. "ipmiMonitor" is a C++ wrapper for "OpenIpmi" interfaces used to communicate with MicroTCA Carrier Hub (MCH) device. It also provides and interface for external tools Command Line (CLI) tool which is used for debugging. Asyn is responsible for EPICS communication layer and synchronising parameters. 

\subsubsection{IPMI Manager IOC Operation}

When IOC is launched, it starts thread for "ipmiMonitor" and separate thread for general refresh processes (it is not possible to read the state of modules, sensors continuously, it takes several seconds to rebuild the internal structure, that is why refresh intervals were defined). Asyn driver task is to create internal parameters for all properties to be monitored via Chanel Access (CA) and to share parameters in PVs.

What is worth mentioning, to make this solution scalable, the dynamic approach was applied for IOC and for operator panels. In case of IOC, it means that there are no defined database file with PVs. It has only templates that are used to generate substitutions files and then PVs. During first launch, IOC detects the crate configuration and generates substitution files depending on available modules and sensors. Thanks to that, there are no redundant PVs from non-existent modules and there is no need to define many template files for each of available module for MTCA.4 crate in ESS. The same approach was introduced to GUI. Using Jython scripts some elements are generated basing on available PVs and their values. We avoided designing user interface for each type of module.

\section{Test and measurements}
\label{sec:test}

\subsection{Cavity Simulator Model Measurements}

\begin{figure}[h!]
\centerline{\includegraphics[width=3.5in]{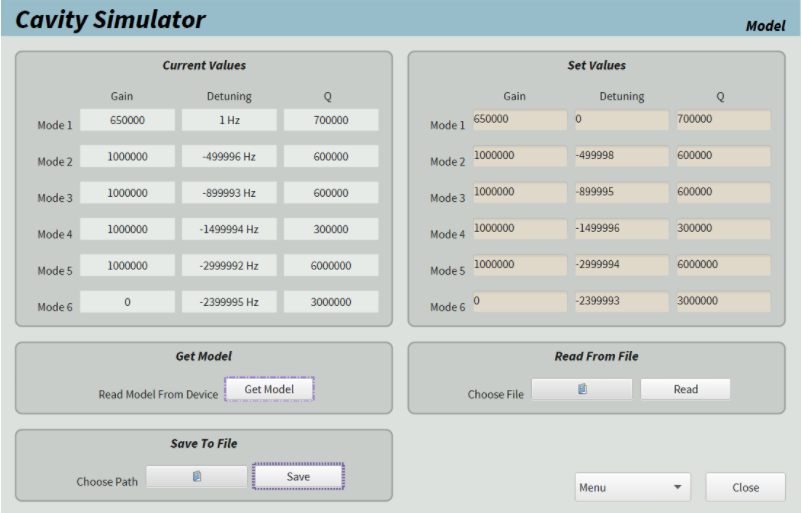}}
\caption{Parameters of Medium-Beta loaded with Cavity Simulator GUI to IOC}
\label{fig8}
\end{figure}

To verify proper functioning of the Cavity Simulator's IOC the measurements with network analyser, connected in the loop with the RF input and the cavity probe, have been carried out. In this configuration, analyser is able to detect network changes caused by response of the model. \figurename~\ref{fig7} shows configuration of the Medium-beta model in the IOC. The applied quality factor, gain, and detuning values results from measurements conducted on the real cavities in 2 K temperature.

\begin{figure}[h!]
\centerline{\includegraphics[width=3.5in]{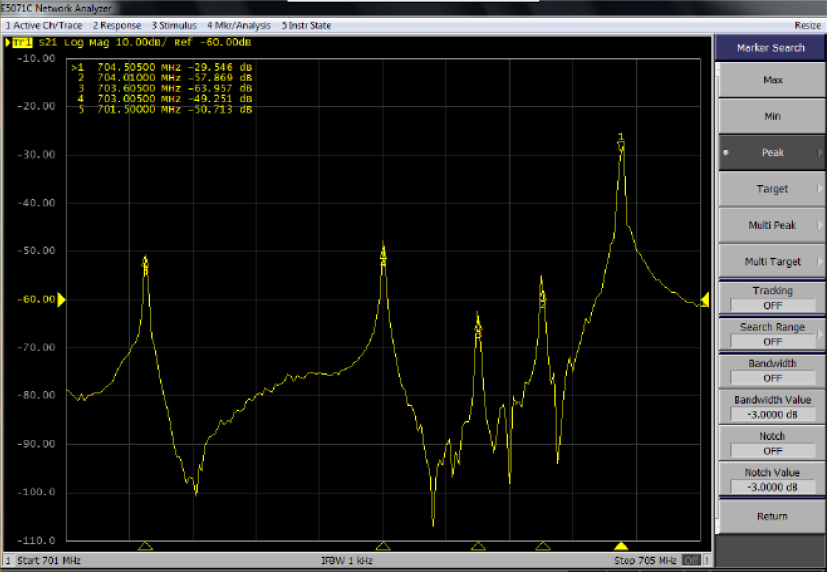}}
\caption{Medium-Beta model response}
\label{fig9}
\end{figure}

\figurename~\ref{fig8} presents the results of the tests with network analyser. The detuning values from the operator panel correspond to distances between modes. For instance, the detuning value of \nth{3} mode is set to 1.5 MHz. The same value can be read from the frequency analyser. Having that, we could confirm that IOC fulfill its role.

\subsection{Piezo Driver Excitation Signal Generation}
 
\begin{figure}[h!]
\centerline{\includegraphics[width=3.5in]{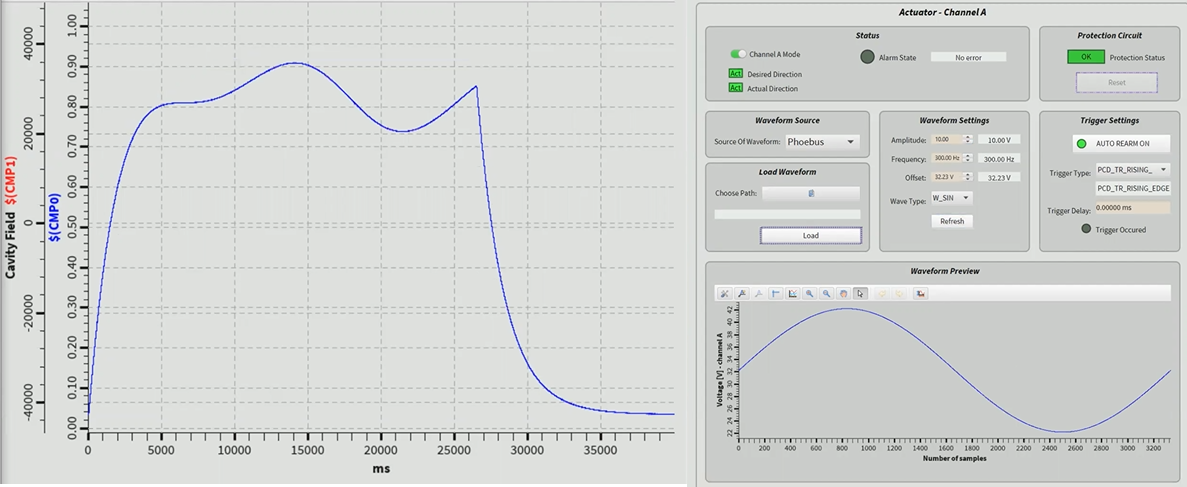}}
\caption{Piezo Driver window and answer of the cavity model for sine excitation signal}
\label{fig10}
\end{figure}

The Piezo Control System is responsible for Lorentz force compensation by introducing physical tuning of the cavity. We have tested E3 module using the Cavity Simulator in order to check whether the excitation signal has an influence on the cavity response. Firstly, the model of the cavity was loaded with Cavity Simulator IOC. Then, LLRF IOC received the expected cavity response, in accordance with Medium-Beta parameters. Next, the Piezo Driver IOC was started and the excitation signal was generated. To emphasize its impact, we chose sine as a signal wave. The results are visible in \figurename~\ref{fig9}. Left side of the figure presents cavity response with noticeable sine shape. Right section shows output signal generated in the IOC. Other tests have been performed with various shapes, amplitudes, offsets, and frequencies. They confirmed that software functions properly. 

\subsection{LO Distribution Tool Functioning}

\begin{figure}[h!]
\centerline{\includegraphics[width=3.5in]{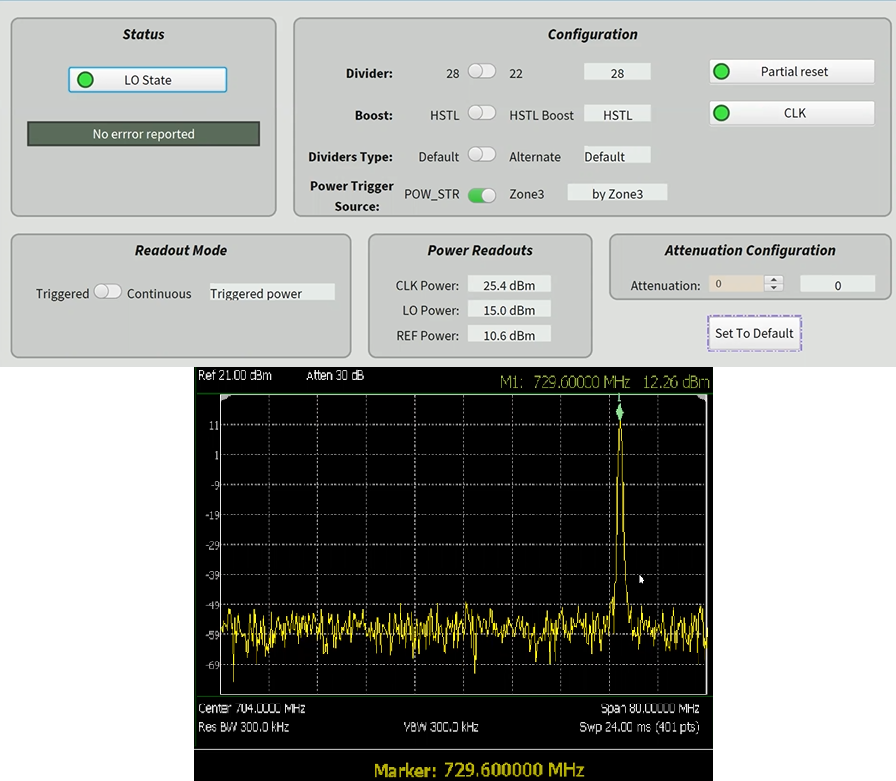}}
\caption{GUI of LO Distribution Tool and result from frequency analyser}
\label{fig11}
\end{figure}

We conducted measurements to verify if LO RTM works as expected and the LO Distribution Tool is able to control it. \figurename~\ref{fig10} shows part of the results. In the operator panel we can see that LO is turned on. That is in accordance with LO power readout which is equal to 15.0 dBm (when it is turned off it is around -17.0 dBm) and with screenshot from frequency analyser. The peak at 729 Mhz is noticeable and it means that LO is working properly. The clock power readout (25.4 dBm) informs that clock is also turned on.

\subsection{IPMI Manager Functional Tests}

\begin{figure}[h!]
\centerline{\includegraphics[width=3.5in]{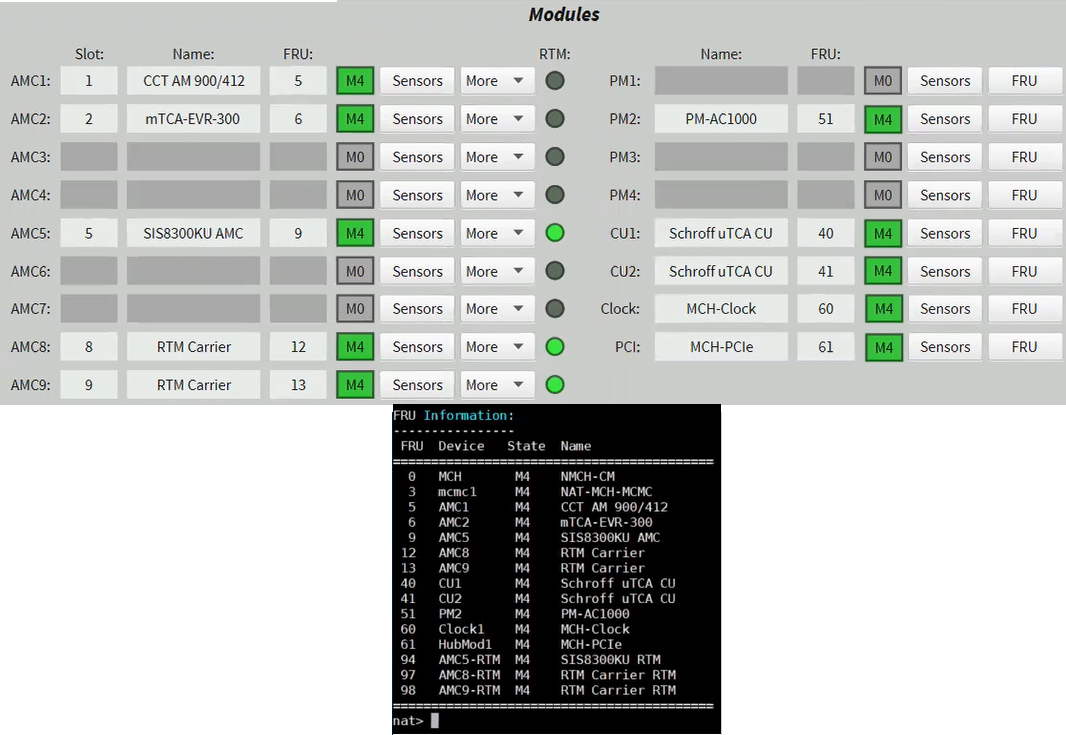}}
\caption{IPMI Manager IOC panel operator and NAT console output}
\label{fig12}
\end{figure}

The only way to validate the IPMI Manager operation is comparison of the results from MCH console output with information from PVs represented in operator panels. 

MicroTCA Carrier Hub (MCH) is a module responsible for management, data switching, and health monitoring for all components of a MicroTCA.4 system. The MCH we use is composed of the base module and numerous daughter cards. It distributes clock signal to all occupied AMC slots. It also features a command line interface using serial port or Telnet protocol. This console can be used to configure parameters of the MCH and to access information about all installed modules (number of sensors, their values, units, and thresholds) \cite{b20}. 

\figurename~\ref{fig11} presents the MCH console output for "show\_fru" command and first level of the GUI where user can check what type of modules are available in the system. We continued with tests of sensor readouts for different modules and e-keying information. In both cases results from MCH console have been entirely covered within IOC. Thresholds of the sensors have been assigned to records alarm fields of the sensor values. That results in border color changes when one of the thresholds is exceeded. We have testes setting thresholds and time intervals. Both control functionalities work in accordance with requirements.

Currently the IPMI Manager is able to monitor and control following properties:
\begin{enumerate}
    \item read all Field Replaceable Unit (FRU) information,
    \item read all sensor values information (units, type etc.),
    \item read sensor values and flags
    \item read and set sensors thresholds,
    \item read hot plug state.
\end{enumerate}
Those functionalities work for AMCs slots, RTMs slots, Cooling Units, (CUs), Power Modules (PMs), Clock and PCIe boards.

\section{Conclusion}
\label{sec:conclusion}
In this paper, the software tools for elliptical cavities LLRF system peripherals management are described. Their proper functioning have been confirmed with functional tests. The Cavity Simulator application emulates the behaviour of cavities model of two types Medium-Beta and High-Beta. It can be used for LLRF control system measurements and evaluation. With Piezo Driver tool, we can set the different type of excitation signal for cavity tuning. Its influence is visible on responses from cavity model. The LO Distribution IOC manages local oscillator signal of accurate frequency and provides possibility of its configuration. The IPMI Manager meet all defined requirements. It allows for reading all necessary information about the state of the crate. Thanks to dynamic creation of databases files, it can be adapted to almost any configuration of MTCA crate. The information from IOC corresponds to external monitoring tool console output. The developed software presents them in the more user-friendly way. 

The development of all modules is in progress to extend their functionalities. We are examining software for finding bugs and for improving their performance. If it comes to the IPMI Manager, currently, tests are being performed on ESS side with various create configurations. Piezo Driver, LO Distribution and Cavity Simulator tools are being integrated with LLRF control system and actively used in this control systems integration and testing process.

\end{document}